\documentclass[aps,prd,11pt,superscriptaddress]{revtex4-1}
\usepackage{amsmath}    
\usepackage{graphicx}
\usepackage{verbatim}
\usepackage{color} 
\usepackage{subfigure}
\usepackage{hyperref}

\begin{document}

\title{Model-independent top quark width measurement using a combination of resonant and non-resonant cross sections.}

\author{Alexey Baskakov}
\affiliation{Skobeltsyn Institute of Nuclear Physics M.V. Lomonosov Moscow State University, Moscow 119991, Russian Federation}
\author{Eduard Boos}
\affiliation{Skobeltsyn Institute of Nuclear Physics M.V. Lomonosov Moscow State University, Moscow 119991, Russian Federation}
\affiliation{Faculty of Physics M.V.Lomonosov Moscow State University Leninskie Gory, Moscow 119991, Russian Federation}
\author{Lev Dudko}
\affiliation{Skobeltsyn Institute of Nuclear Physics M.V. Lomonosov Moscow State University, Moscow 119991, Russian Federation}

\begin{abstract}
Though the top quark was discovered more than twenty years ago, measurement of its width is still challenging task. Most measurements either have rather low precision or they are made under the assumption of the SM top quark interactions. We consider model-independent parametrization of the top quark width and provide estimations on achievable accuracy using a combination of fiducial cross sections in double-resonant, single-resonant and non-resonant regions.
\end{abstract}

\maketitle

\section{Introduction}
\label{intro}
The top quark is the heaviest known elementary particle. This fact makes it along with the Higgs boson the most promising window to physics Beyond the Standard Model (BSM). Measurements of the top quark properties and parameters are crucial for testing deviations from the Standard Model (SM) predictions.
While the top quark mass was  measured directly with an accuracy at the percentage level~\cite{Patrignani:2016xqp}
the direct measurements of the top quark width give much worse precision of about 50\% mainly because of low experimental resolution~\cite{Aaltonen:2013kna}. Recent results of direct measurements of the width presented by CMS and ATLAS collaborations are of $0.6< \Gamma_{\rm t} < 2.5$ GeV~\cite{CMS:2016hdd} and $\Gamma_{\rm t} = 1.76 \pm 0.33 (\rm stat.) ^{+0.79}_{-0.68} (\rm syst.)$ GeV~\cite{Aaboud:2017uqq}.
The indirect top quark width measurements have reached an accuracy of about ten percent~\cite{Abazov:2012vd, Khachatryan:2014nda}. However, the top quark width was measured indirectly only under certain SM assumptions, in particular assuming only the SM decay modes.

The accuracy of the direct top quark width measurement is expected to be improved by the analysis of the b-charge asymmetry with $\rm W^+,b$, $\rm W^-,\bar b$ final states for the s-channel top, anti-top quark resonant contribution and  with $\rm W^-,b$, $\rm W^+,\bar b$ final states for non-resonant top quark contribution ~\cite{Giardino:2017hva}.  

In this paper, we discuss another method of setting model-independent limits on the top quark width in completely gauge invariant way by fitting fiducial cross sections of $\rm W^+bW^-\bar b$ production in certain phase space regions called double-resonant, single-resonant, and non-resonant.  A similar method for the case of $\rm e^+e^-$ collisions has been discussed in~\cite{Liebler:2015ipp}. The idea of the method was illustrated 
on a simple $2\to 3$ example for the process $\rm gg\rightarrow tW^-\bar{b}$ in~\cite{Baskakov:2017jhb}.
This work is a generalization of that study.

The idea of the width measurement from the comparison of rates in the on-shell and off-shell phase space regions was previously proposed for the Higgs boson~\cite{Kauer:2012hd,Caola:2013yja}. 
In corresponding measurements, the Higgs boson width is extracted from $\rm pp \rightarrow ZZ$ production above 
the $ZZ$ threshold and from $\rm pp \rightarrow H \rightarrow ZZ^*$ production below the threshold in the $\rm ZZ^*$ mass region close the Higgs boson mass. This approach can not be directly applied to the top quark. The Higgs boson is a substantially narrower resonance than the top quark. This fact allows calculating separately amplitudes for $\rm pp \rightarrow ZZ$ and $\rm pp \rightarrow H \rightarrow ZZ^*$ processes in a gauge invariant way. In case of the off-shell top quark production with its subsequent decay to $Wb$ one cannot make calculations of diagrams involving the top quark pair and the single top separately in a gauge invariant way. Therefore we perform the computation of the complete gauge invariant set of diagrams and investigate a sensitivity of fiducial cross sections to deviations from the SM caused by the top quark width and related Wtb coupling. This approach enables one to put model-independent and fully gauge invariant constraints on the top quark width. 

\section{Numerical simulation}
\subsection{The top quark width parametrization}
\label{sec21}

The total top quark width can be parametrize as follows
\begin{equation}\label{eq3}
\Gamma_{\rm t} = \xi^2 \cdot \Gamma^{SM}_{\rm t} + \Delta ,
\end{equation}
reflecting that the top quark width may differ from its SM value either by a modification of the Wtb coupling (e.g. see~\cite{Alwall:2006bx}) or by a presence of additional non-SM decay modes (e.g. see~\cite{Han:1996ep, Larios:2006pb}). 
In Eq.~\ref{eq3} the parameter $\xi$ simultaneously changes the top quark width and rescales the $Wtb$ coupling.  
The parameter $\Delta$ affects the top quark width only. One should note that the production cross section times branching ratio remains unchanged with the variation of the parameter $\xi$ in case of $\Delta = 0$.
It is useful to parametrize the deviation $\Delta$ also in terms of the SM top quark width as $\Delta=\delta \cdot\Gamma^{SM}_{\rm t}$. 

In the SM $\xi=1$ and $\delta=0$. In order to study deviations of the top quark width from its SM value  it is more convenient to have two parameters equal to zero in the SM and introduce the parameter $\epsilon$ instead of $\xi$ as follows
\begin{equation}\label{eq4}
\epsilon = \xi^2 - 1.
\end{equation}

It should be stressed, that the parameters $\epsilon$ (or $\xi$) and $\delta$ have a different origin, affect the matrix element in a different way, and therefore cannot be combined in a single parameter. 

\subsection{Numerical results for \texorpdfstring{$\rm pp\rightarrow W^+W^-b\bar{b}$}{pp->WWbb}}
\label{sec22}

To illustrate expediency of the entered parametrization we consider a complete tree-level set of Feynman diagrams for the process \textit{}$\rm pp\rightarrow W^+W^-b\bar{b}$, where both top quarks are off-shell. As is well known, the main contribution comes from the gluon fusion subprocess~\cite{Moch:2008qy}, however, we take into account the contributions from all partonic subprocesses. The CompHEP generator~\cite{Boos:2004kh} with MSTW2008 PDF~\cite{Martin:2009iq} is used for the calculation. The computations are performed for a certain value of the top quark mass, for a definiteness it was taken to be $m_{\rm t}= 172.5$ GeV, and for various values of the top quark width with the corresponding rescaling of the Wtb coupling. In numerical computations, the LO value of the top quark width was taken to be $\Gamma^{SM}_{\rm t} = 1.49$ GeV. Fixed scale of $m_t$ was used. The change of the scale in the range $m_t/2 - 2 \times m_t$ does not make any practical influence on the patterns presented below. Calculations were carried out in 4-flavor scheme with massive b-quark.

Hadronization and fragmentation effects, as well as backgrounds impact, are postponed to the next more realistic analysis, not to distract from the main idea of this research.  Realistic estimations of these effects are included in systematic uncertainty estimations.

The NLO QCD corrections for the process $\rm pp\rightarrow W^+W^-b\bar{b}$ were computed~\cite{Denner:2010jp} showing an impact on various kinematic distributions and making results more stable with respect to the QCD scale variation. The NLO corrections to the complete $2\to 6$ process involving off-shell W bosons were calculated~\cite{Bevilacqua:2010qb, Denner:2012yc, Denner:2016jyo, Denner:2017kzu} and the k-factor for $13$ TeV LHC energy was found to be 1.16. At this stage of our analysis, which aims to show the main effect caused by the width change, the complete leading order contributions have been taken into account, and the impact of the NLO corrections has been included in the assumed systematic uncertainties, as will be explained below. 
 
The boundaries of fiducial double-resonant (DR), single-resonant (SR) and non-resonant (NR) regions are expressed in terms of the SM value of the top quark width in the following way.
 
Double-resonant region (DR),
\begin{equation}
\label{eq1}
\Big(m_{\rm t} - n\cdot\Gamma_{\rm t}^{SM}  \leq M_{\rm W^-\bar{b}} \leq m_{\rm t} + n\cdot\Gamma_{\rm t}^{SM}\Big)\quad and \quad \Big(m_{\rm t}- n\cdot\Gamma_{\rm t}^{SM}  \leq M_{\rm W^+b} \leq m_{\rm t} + n\cdot\Gamma_{\rm t}^{SM}\Big)
\end{equation}

Single-resonant region (SR),
\begin{equation*}
\begin{split}
\Big(m_{\rm t} - n\cdot\Gamma_{\rm t}^{SM}  \leq M_{\rm W^-\bar{b}} \leq m_{\rm t} + n\cdot\Gamma_{\rm t}^{SM}\Big)\: &and\:
\Big(M_{\rm W^+b} \leq m_{\rm t} - k\cdot\Gamma_{\rm t}^{SM}  \: or  \:   m_{\rm t} + k\cdot\Gamma_{\rm t}^{SM} \leq M_{\rm W^+b}\Big)
\\ &or \\
\Big(m_{\rm t}- n\cdot\Gamma_{\rm t}^{SM}  \leq M_{\rm W^+b} \leq m_{\rm t} + n\cdot\Gamma_{\rm t}^{SM}\Big)\:&and \:
\Big(M_{\rm W^-\bar{b}}\leq m_{\rm t} - k\cdot\Gamma_{\rm t}^{SM} \: or  \:  m_{\rm t} + k\cdot\Gamma_{\rm t}^{SM} \leq M_{\rm W^-\bar{b}}\Big)
\end{split}
\end{equation*}

Non-resonant region (NR). 
\begin{equation*}
\begin{split}
\Big(M_{\rm W^-\bar{b}}\leq m_{\rm t} - k\cdot\Gamma_{\rm t}^{SM} \quad &or  \quad  m_{\rm t} + k\cdot\Gamma_{\rm t}^{SM} \leq M_{\rm W^-\bar{b}}\Big) \: \\&and\\
\Big(M_{\rm W^+b}\leq m_{\rm t} - k\cdot\Gamma_{\rm t}^{SM}  \quad &or  \quad   m_{\rm t} + k\cdot\Gamma_{\rm t}^{SM} \leq M_{\rm W^+b}\Big)
\end{split}
\end{equation*}

Here $M_{\rm W^+b}$ and $M_{\rm W^-\bar{b}}$ are the invariant masses, $n$ and $k$ are integer numbers with obvious requirement $n \leq k$ to have no overlapping regions. 

Current experimental data~\cite{Abazov:2012vd, Khachatryan:2014nda, Aaltonen:2013kna, CMS:2016hdd, Aaboud:2017uqq} indicate that deviations from the SM for the top quark width should be small. In order not to contradict with this we will study dependencies of fiducial cross sections from two small parameters $\epsilon$ and $\delta$.

\begin{figure}%
    \centering
    \subfigure[~DR region]{{\includegraphics[width=6cm]{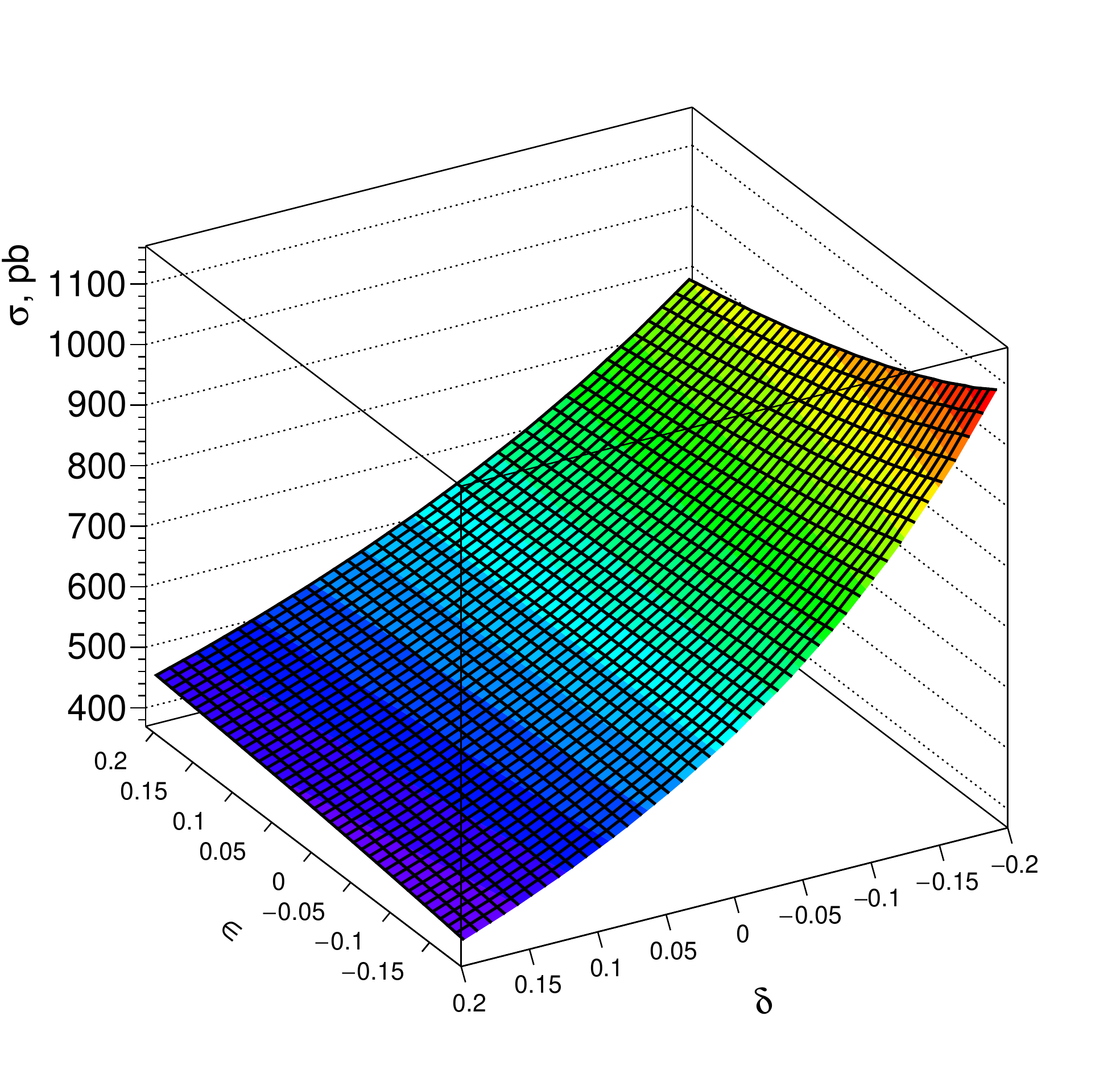} }}%
		\qquad
    \subfigure[~SR region]{{\includegraphics[width=6cm]{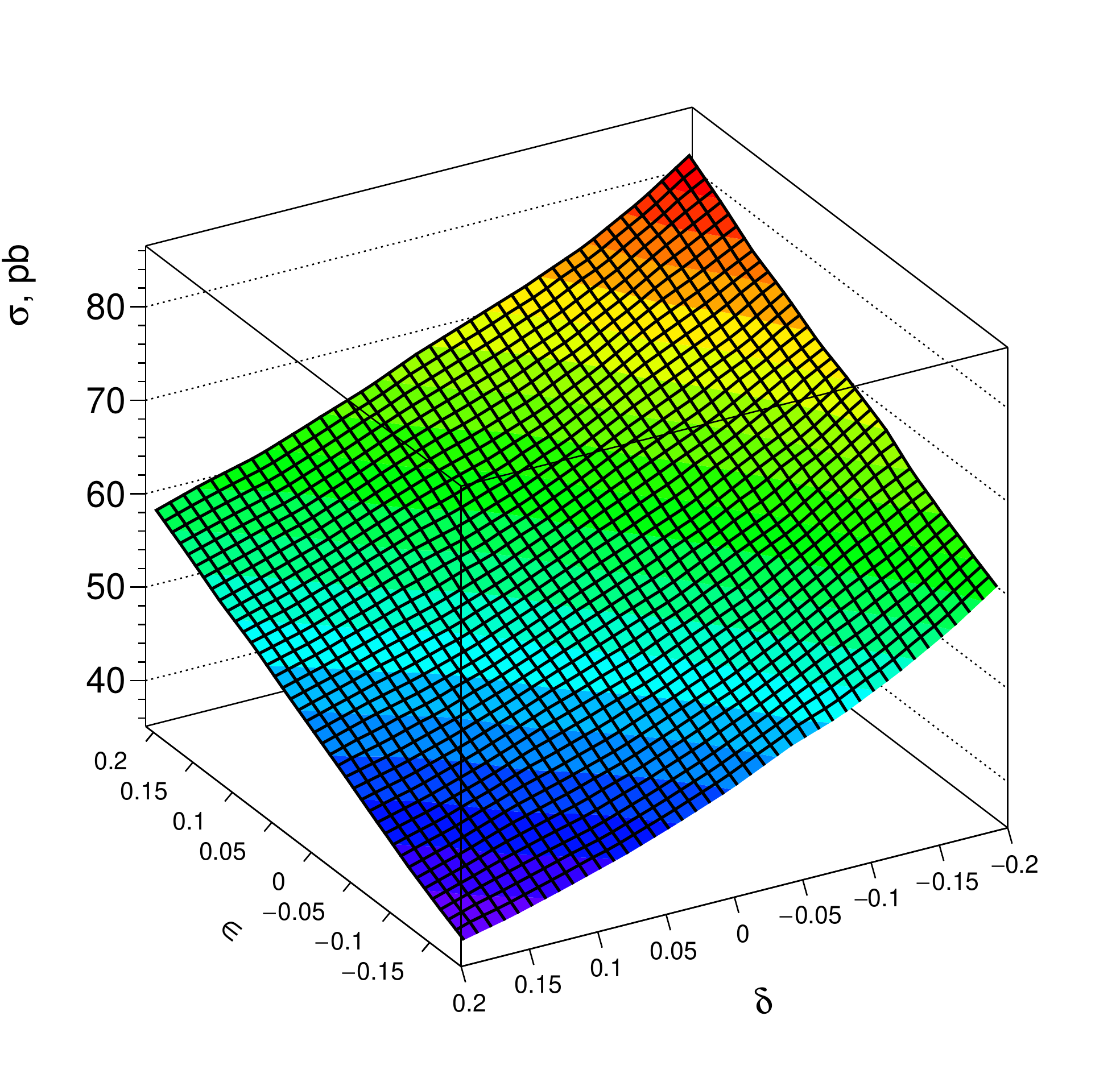} }}%
    \qquad
    \subfigure[~NR region]{{\includegraphics[width=6cm]{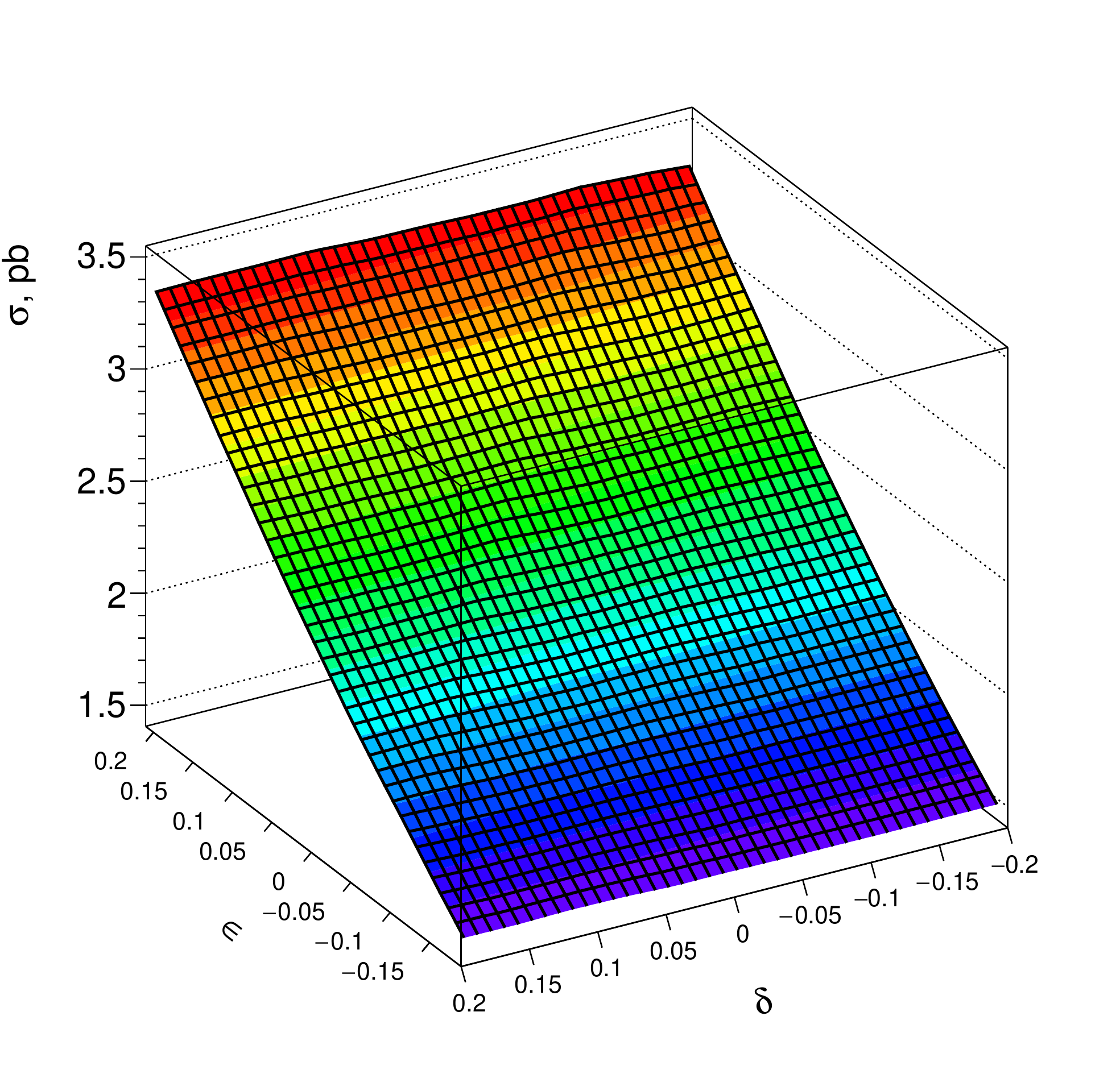} }}%
    \caption{Fiducial cross section dependencies on $\epsilon$ and $\delta$ parameters for the $14$ TeV collision energy, $n=k=15$.}%
    \label{fig:CS_14_TeV}%
\end{figure}

\begin{figure} [h]%
    \centering
    \subfigure[~DR region]{{\includegraphics[width=6cm]{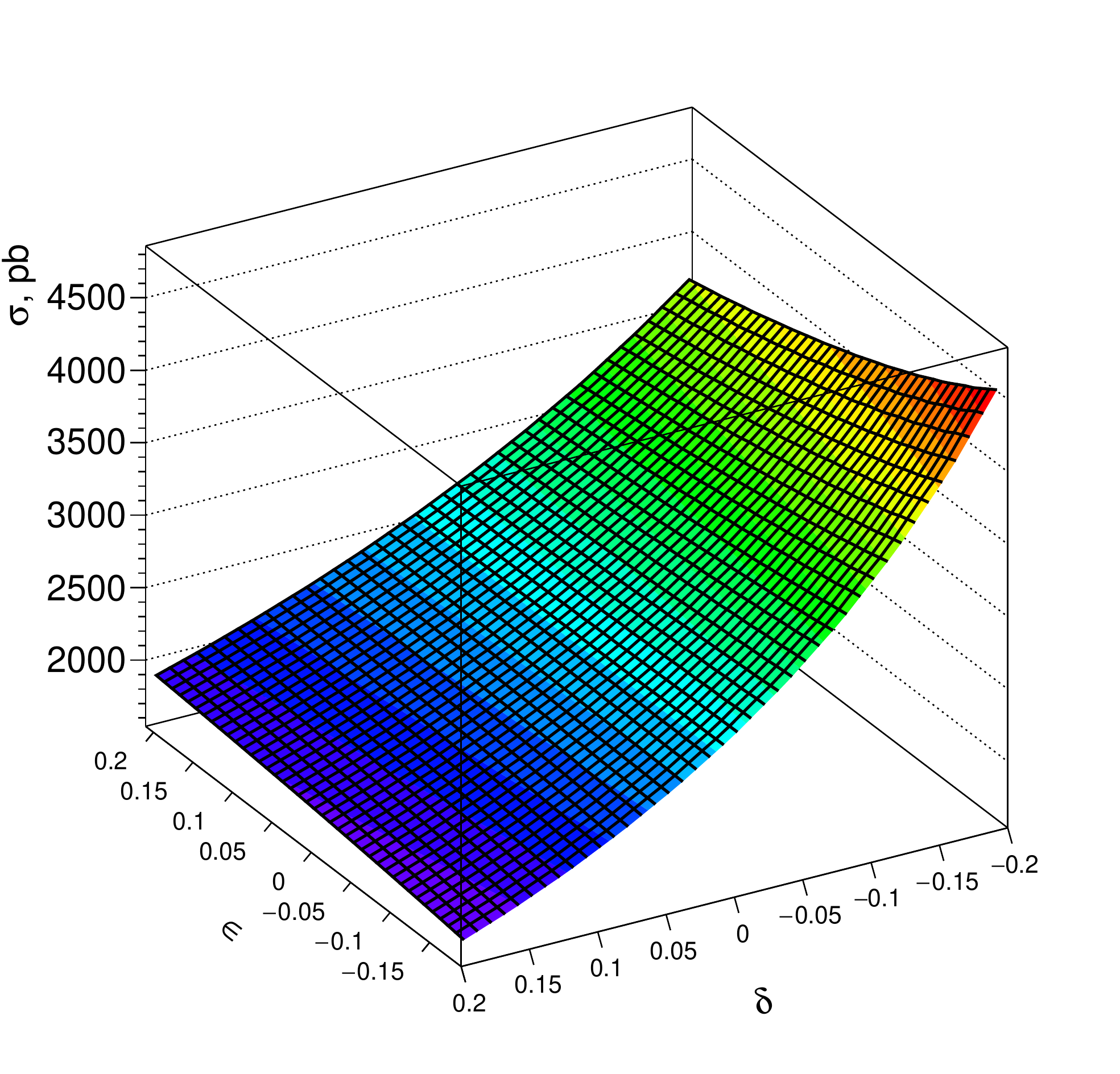} }}%
		\qquad
    \subfigure[~SR region]{{\includegraphics[width=6cm]{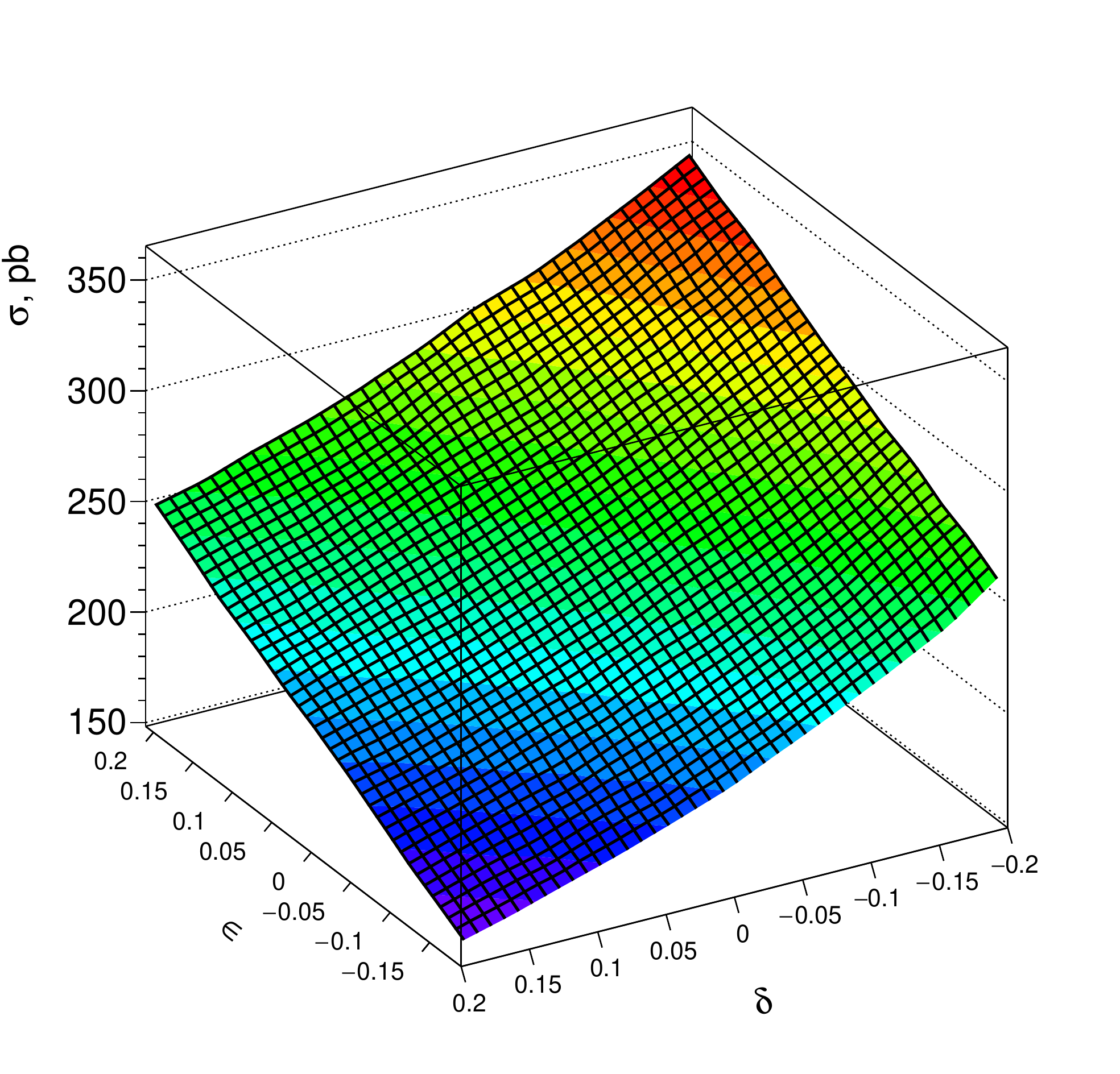} }}%
    \qquad
    \subfigure[~NR region]{{\includegraphics[width=6cm]{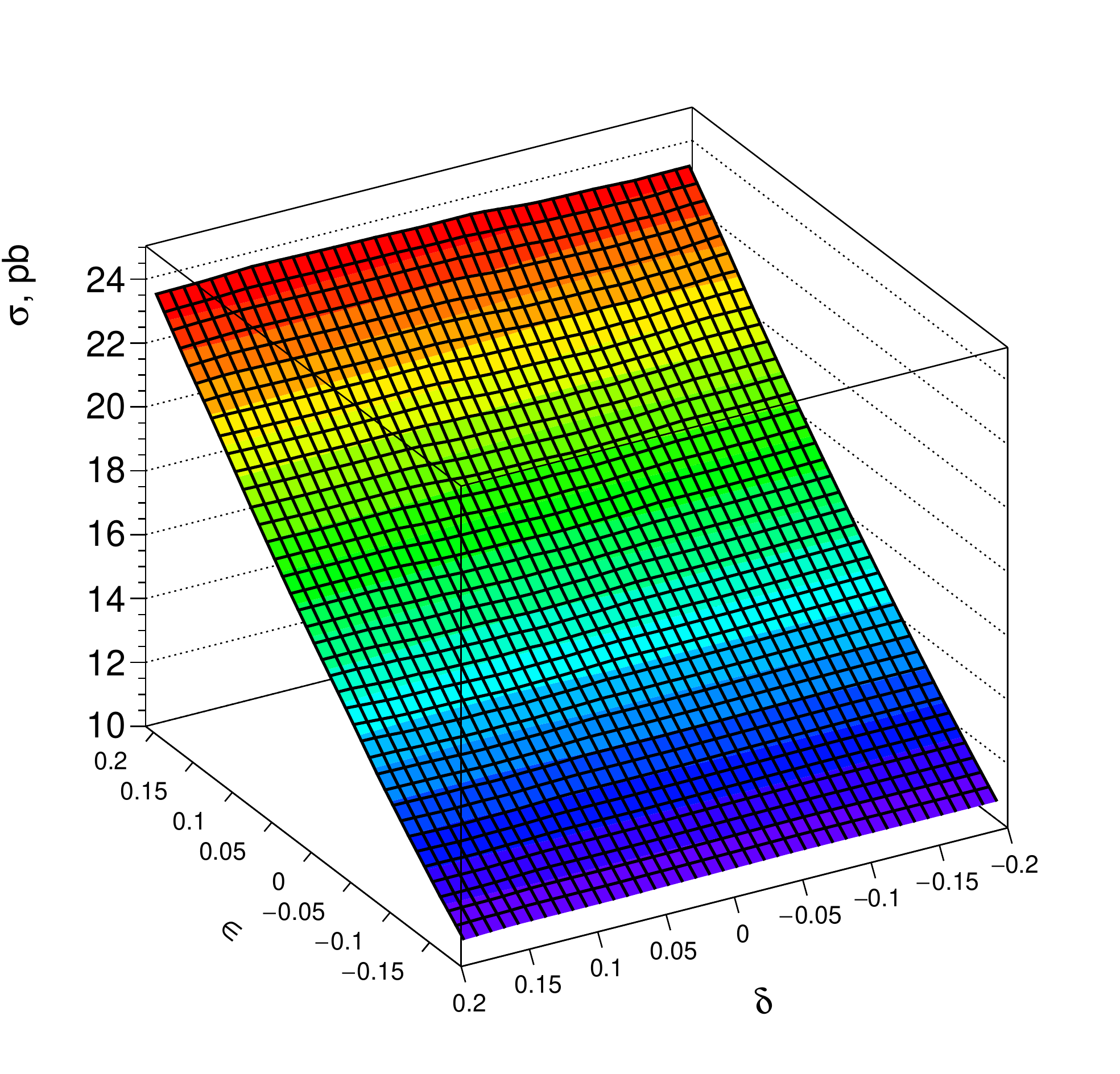} }}%
    \caption{Fiducial cross section dependencies on $\epsilon$ and $\delta$ parameters for the $28$ TeV collision energy, $n=k=15$.}%
    \label{fig:CS_28_TeV}%
\end{figure}

\begin{figure} [h]%
    \centering
    \subfigure[~DR region]{{\includegraphics[width=6cm]{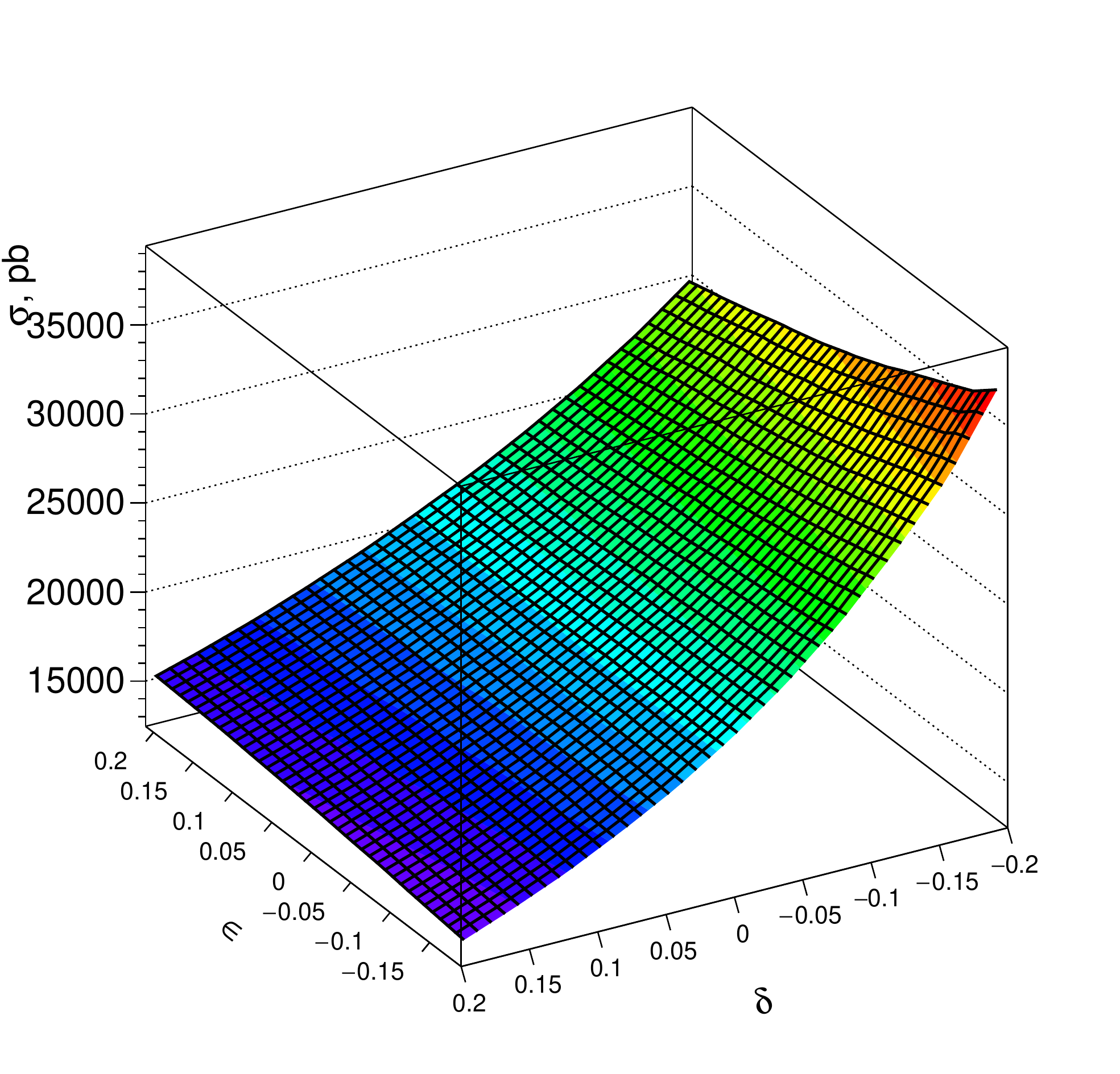} }}%
		\qquad
    \subfigure[~SR region]{{\includegraphics[width=6cm]{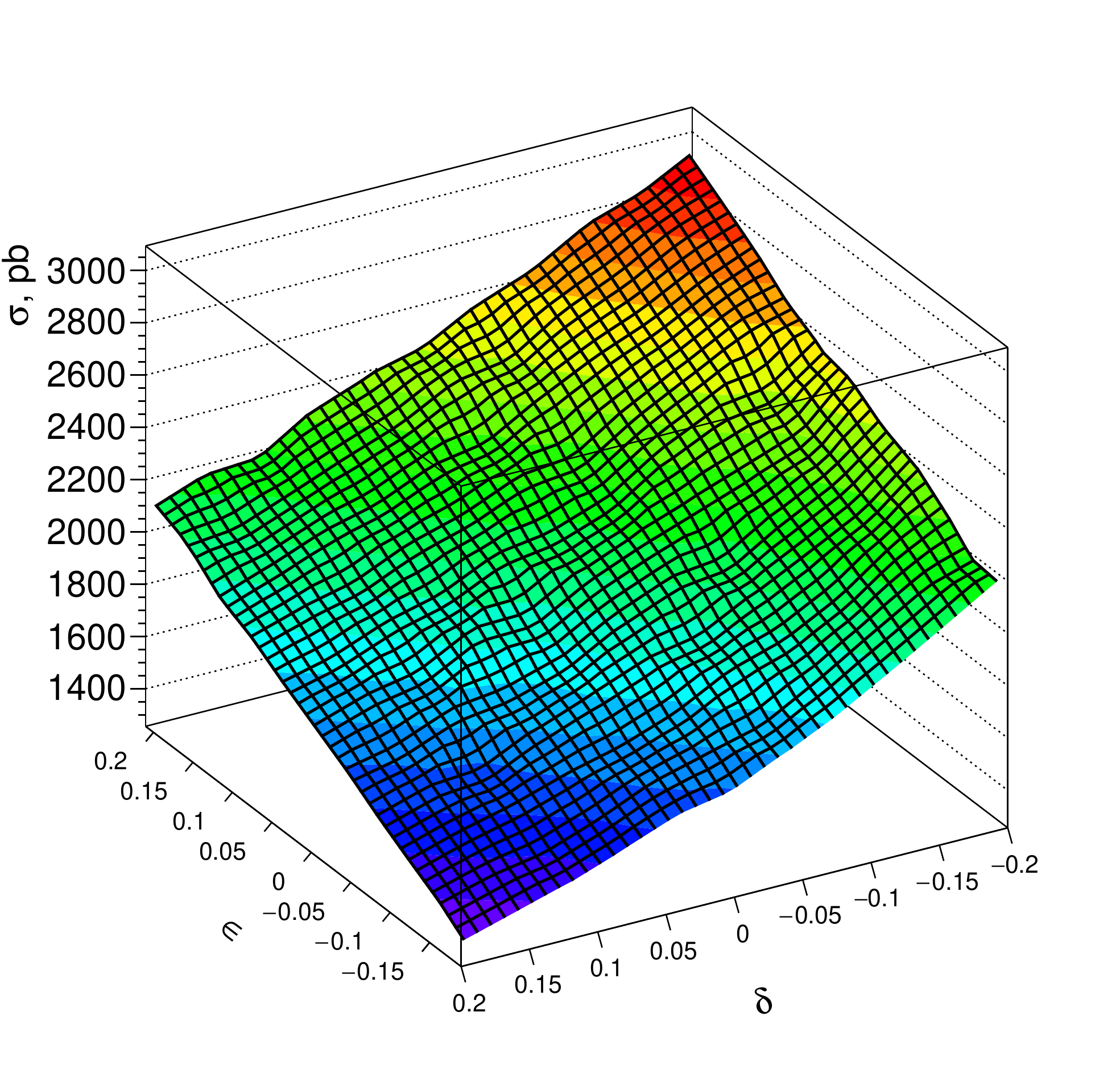} }}%
    \qquad
    \subfigure[~NR region]{{\includegraphics[width=6cm]{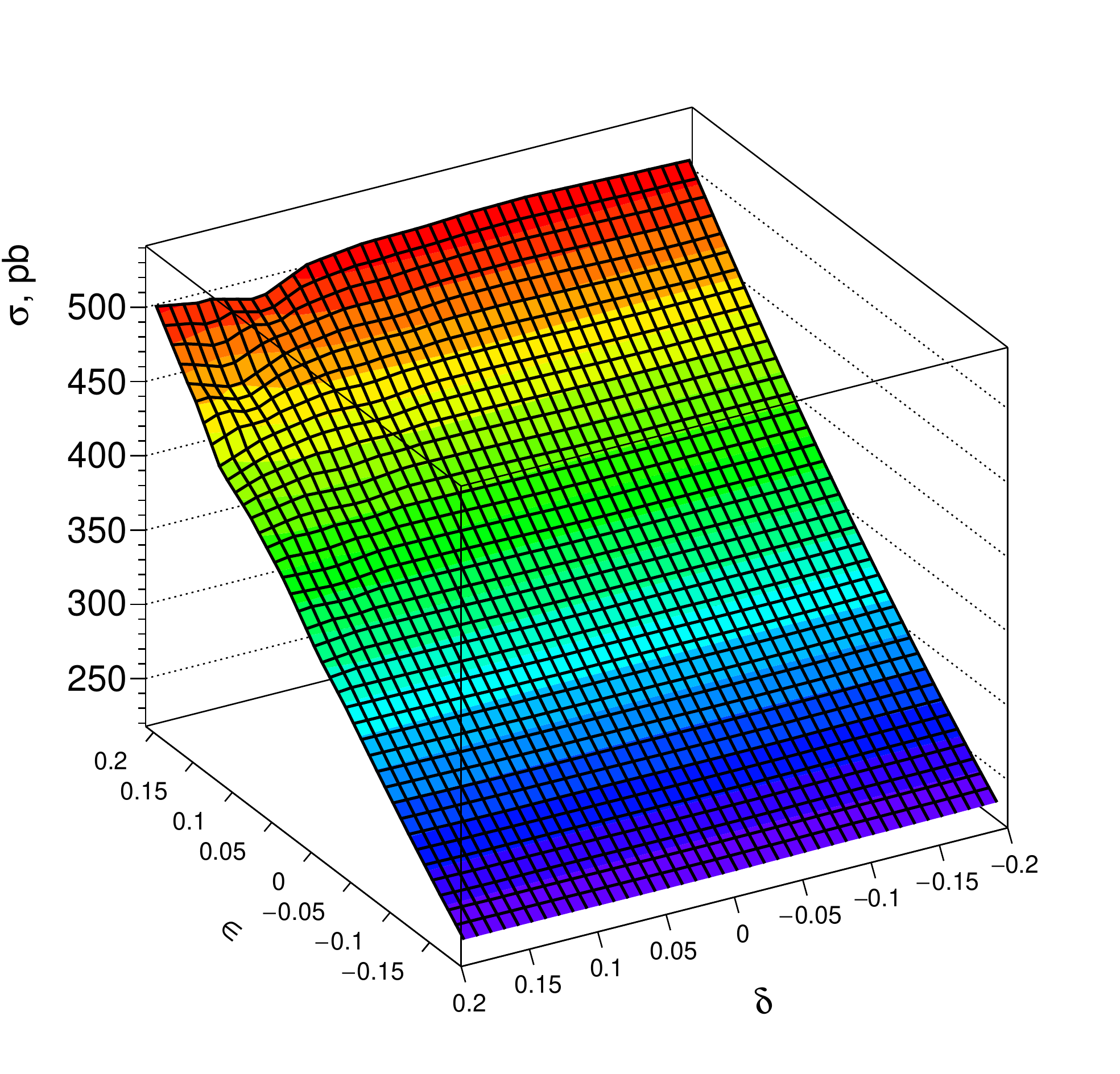} }}%
    \caption{Fiducial cross section dependencies on $\epsilon$ and $\delta$ parameters for the $100$ TeV collision energy, $n=k=15$.}%
    \label{fig:CS_100_TeV}%
\end{figure}

As demonstrated in~\cite{Baskakov:2017jhb}, it is reasonable to select integer parameters $n$ and $k$ in the interval from 10 to  20 for boundaries between the resonant and non-resonant regions, see Eq.~\ref{eq1}.
For definiteness, we take the values $n=k=15$ when $98\%$ of the Breit-Wigner integral concentrated around the pole position~\cite{Kauer:2001sp}.  
Calculation results for the fiducial cross sections at $14$ TeV collision energy in the DR, SR, and NR regions defined above as a function of $\epsilon$ and $\delta$ parameters are shown in Fig.~\ref{fig:CS_14_TeV}. For collision energies of $28$ TeV and $100$ TeV the total rates are substantially higher but the surface shapes are very similar to the case of $14$ TeV, the plots are shown in Figs.~\ref{fig:CS_28_TeV},~\ref{fig:CS_100_TeV}.

As one can see, the surfaces of the three regions have significantly different shapes. The cross-section in the DR region is practically insensitive to simultaneous changes in the top quark coupling and width by the $\epsilon$ parameter since its effect disappears when the amplitude numerator and the denominator are changed simultaneously. 

At the same time, the parameter $\delta$ affects only the amplitude denominator and leads in the  DR region to an inverse polynomial quadratic dependence of the cross section on it. These properties can be easily understood from the Breit-Wigner resonant behavior. In contrast, the cross section in the non-resonant region practically does not depend on the top quark width and therefore depends very weakly on the parameter $\delta$. In turn, the cross section in the NR region depends quadratically on the parameter $\epsilon$ via the coupling constant in the amplitude numerator. One resonance region combines dependence on both types of parameters.
 
The fiducial cross sections in DR, SR, and NR regions are significantly different. The rate in the DR region exceeds by about one order of magnitude the rate in the SR region and by two orders of magnitude the rate in the NR region. The boundaries variation within $10 \div 20$ SM top quark width does not have a significant impact on the cross section rate. NR region has the best sensitivity to the $\epsilon$ parameter but the smallest rate. The DR region has the sensitivity mostly to the direct width modification by the $\delta$ parameter. This fact makes it possible to estimate the top quark width, through the fiducial cross section measurement in the corresponding regions.

Precision measurements of the fiducial cross sections of the top quark production play a crucial role. However in the experimental analyzes $\rm pp\rightarrow W^+W^-b\bar{b}$ process cannot be accessed directly, as W gauge bosons are reconstructed from leptonic or semi-leptonic decays. A decay of the W gauge boson provide signal smearing, the uncertainty of the choice of the four-momentum component of the neutrino and combinatorial factor. Hadronization and fragmentation effects provide additional signal smearing, while standard selection cuts caused by detector geometry and resolution, and b-tagging efficiency decrease the measured rate. It was checked that main influence caused by kinematic cuts is decreasing in the events rate, while main declared dependencies on the  width modifying parameters remains mostly unchanged. This should be taken into account in the experimental analysis, since soft area cuts has a significantly greater influence on the NR region cross section than the DR one. However, the purpose of this paper is to demonstrate the effect of the top quark width parametrization influence. Also, background processes should be taken into account. Accounting of all the listed effects is beyond the scope of this paper and is planned to be implemented within specialized research.
Experimental analysis precision is limited by systematic uncertainties of the jet energy scale, b-tagging efficiency, and luminosity \cite{Sirunyan:2017uhy, Aaboud:2017qyi}. In this regard, we ask the following question. How accurately would we be able to limit the top quark width, if we knew the corresponding fiducial sections in DR, SR and NR regions with a given accuracy?

\subsection{Fitting procedure}
\label{sec23}
To address the top quark width we provide fitting procedure with the standard $\chi^2$ method. 
\begin{equation}\label{eq5}
\chi^2(\sigma) = \left(\frac{\sigma^{SM} - \sigma }{\Delta\sigma}\right)^2,
\end{equation}
where $\sigma^{SM}$ corresponds to SM cross sections ($\epsilon=\delta=0$), $\sigma $ is the cross section with modified top quark width.
$\Delta\sigma = \sigma_{SM}\sqrt{\Delta_{stat}^2+\Delta_{sys}^2}$ represents intended experimental precision. Statistical uncertainties are below percent level today and will decrease further on with high luminosity updates \cite{Apollinari:2015bam}. Evaluable luminosity is taken as $30$ and $300\ fb^{-1}$  for $14$ TeV, $300\ fb^{-1}$ for $28$ TeV and $3000\ fb^{-1}$ for $100$ TeV. It is natural to expect improvement of experimental techniques and decrease of systematic uncertainty as well. The present measurements of the top quark width demonstrate about 10\%~\cite{Khachatryan:2014nda} and 50\%~\cite{Aaboud:2017uqq} uncertainty depending on the analysis method. The present uncertainty of the cross section measurement of tW process is about 10\%~\cite{Sirunyan:2018lcp}. We assume the feasible experimental accuracy of  $10\%$,  $8\%$ and $5\%$ for $14$, $28$ and $100$ TeV collision energies, respectively, including theoretical uncertainties at NLO, NNLO and, possibly, higher level by the time when new high energy machines will be realized. 

As it was mentioned above, the top quark mass is taken $m_{\rm t}= 172.5$ GeV, the LO value of the top quark width is taken to be $\Gamma^{SM}_{\rm t} = 1.49$ GeV.

\begin{figure}%
    \centering
    \subfigure[~DR region]{{\includegraphics[width=6cm]{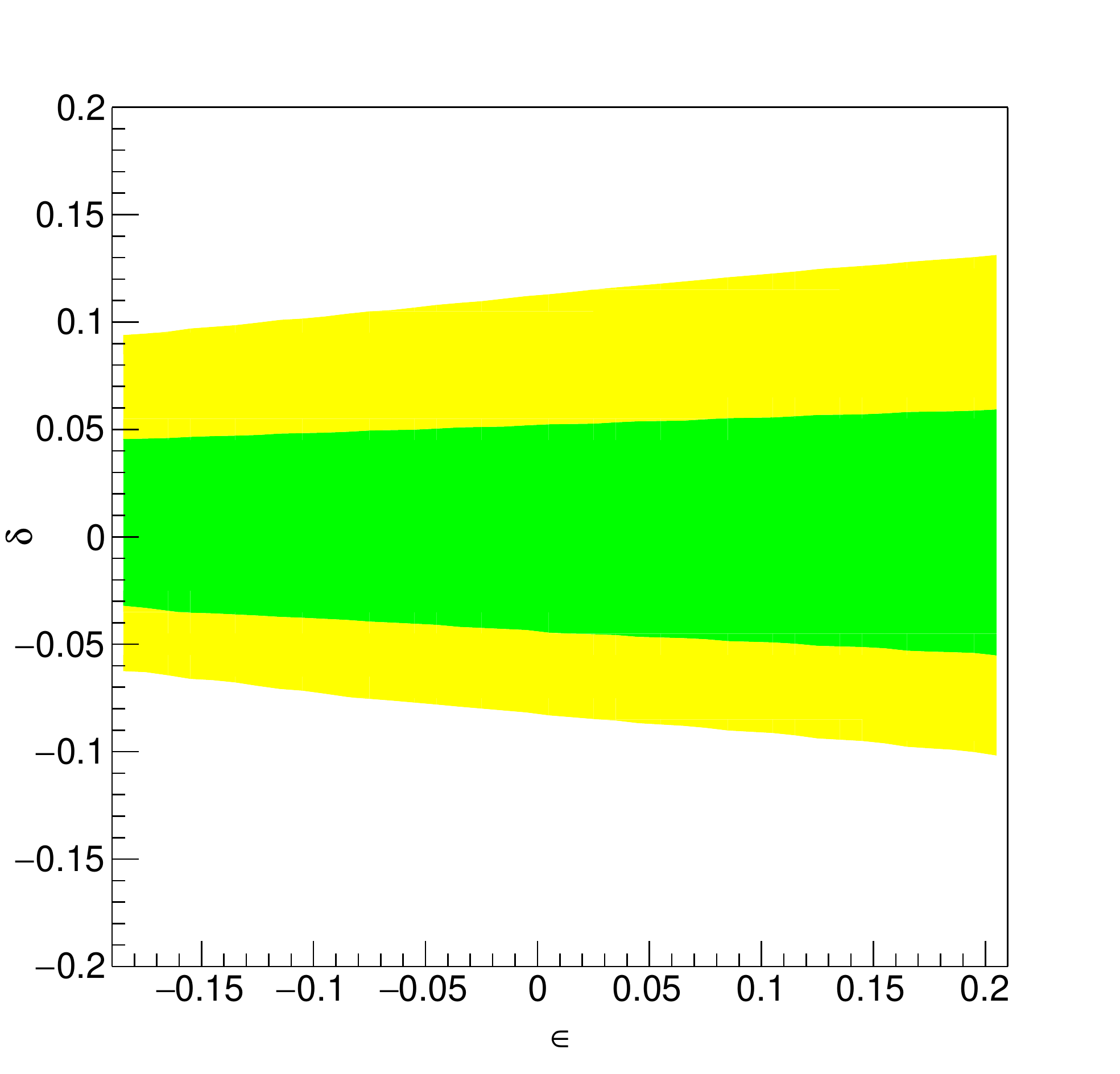} }}%
		\qquad
    \subfigure[~SR region]{{\includegraphics[width=6cm]{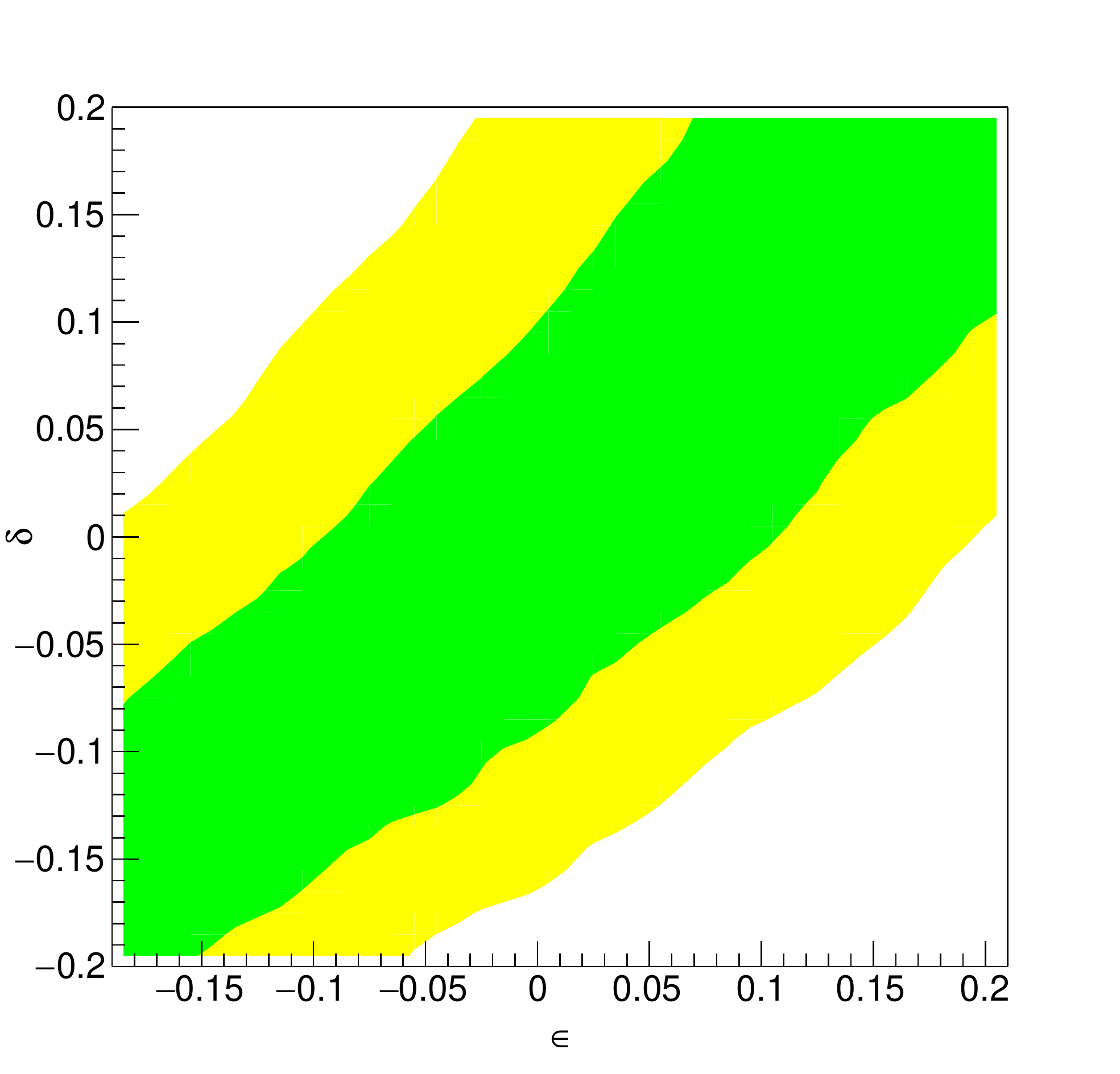} }}%
    \qquad
    \subfigure[~NR region]{{\includegraphics[width=6cm]{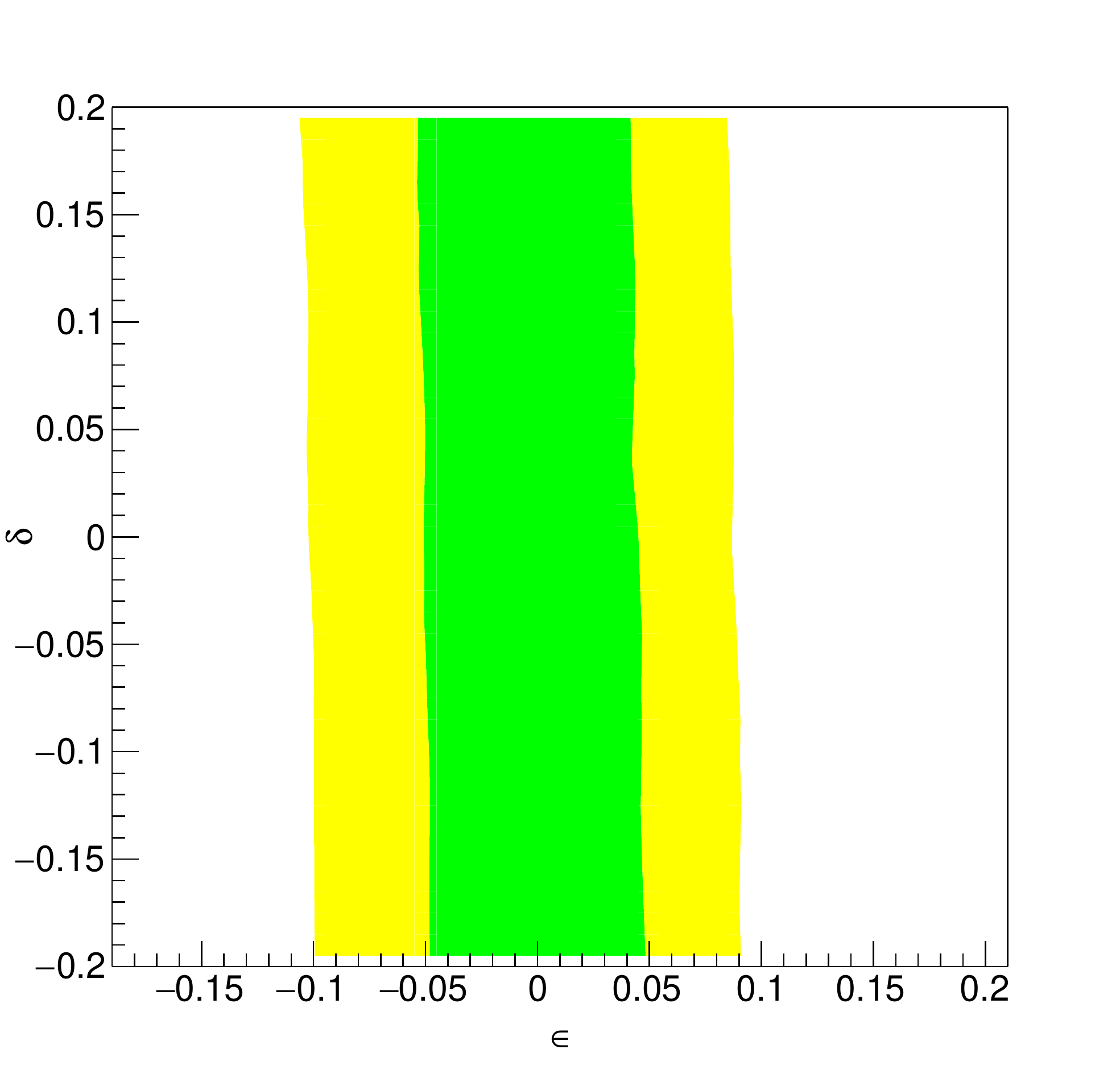} }}%
    \caption{Constraints on   $\epsilon $ and $\delta$ parameters for $14$ TeV collision energy, $n=k=15$ boundary for DR, SR and NR regions. Green and yellow areas correspond to exclusion limits  at $68\%$ and $95\%$ CL on $\epsilon $ and $\delta$ for the CS measured with $10\%$ uncertainty.}%
    \label{fig:14_TeV_limits}%
\end{figure}

\begin{figure}%
    \centering
    \subfigure[~DR region]{{\includegraphics[width=6cm]{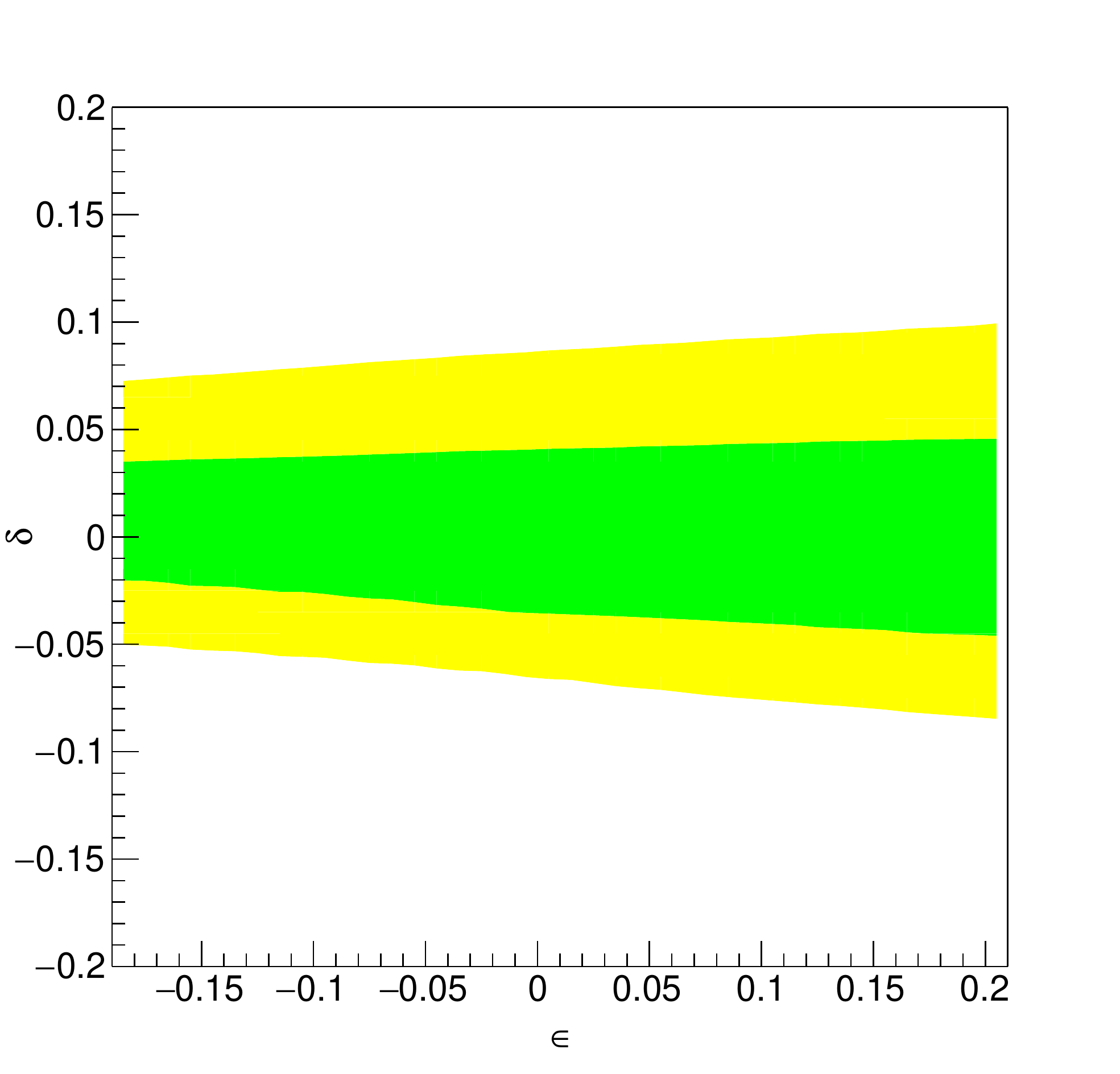} }}%
		\qquad
    \subfigure[~SR region]{{\includegraphics[width=6cm]{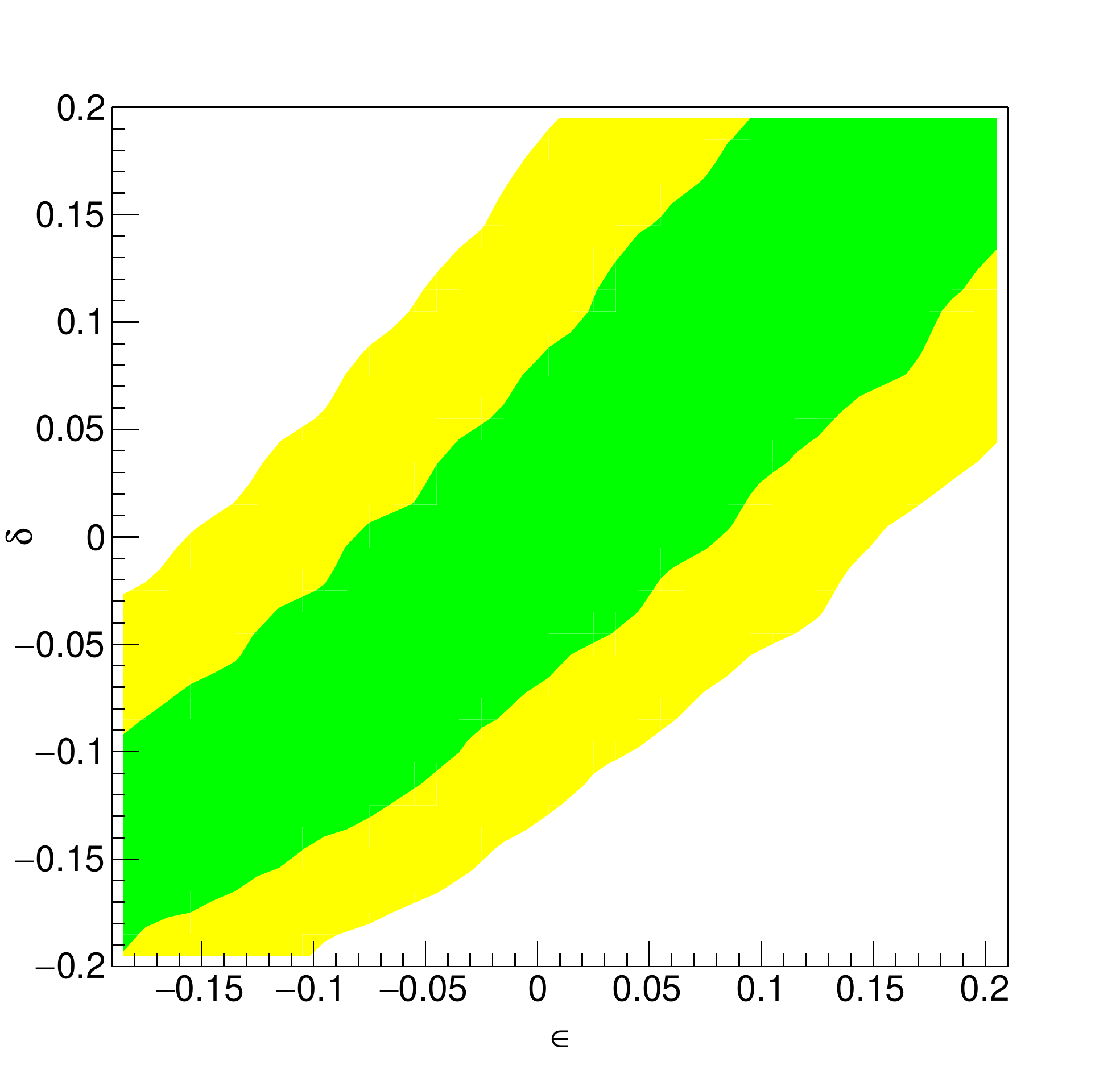} }}%
    \qquad
    \subfigure[~NR region]{{\includegraphics[width=6cm]{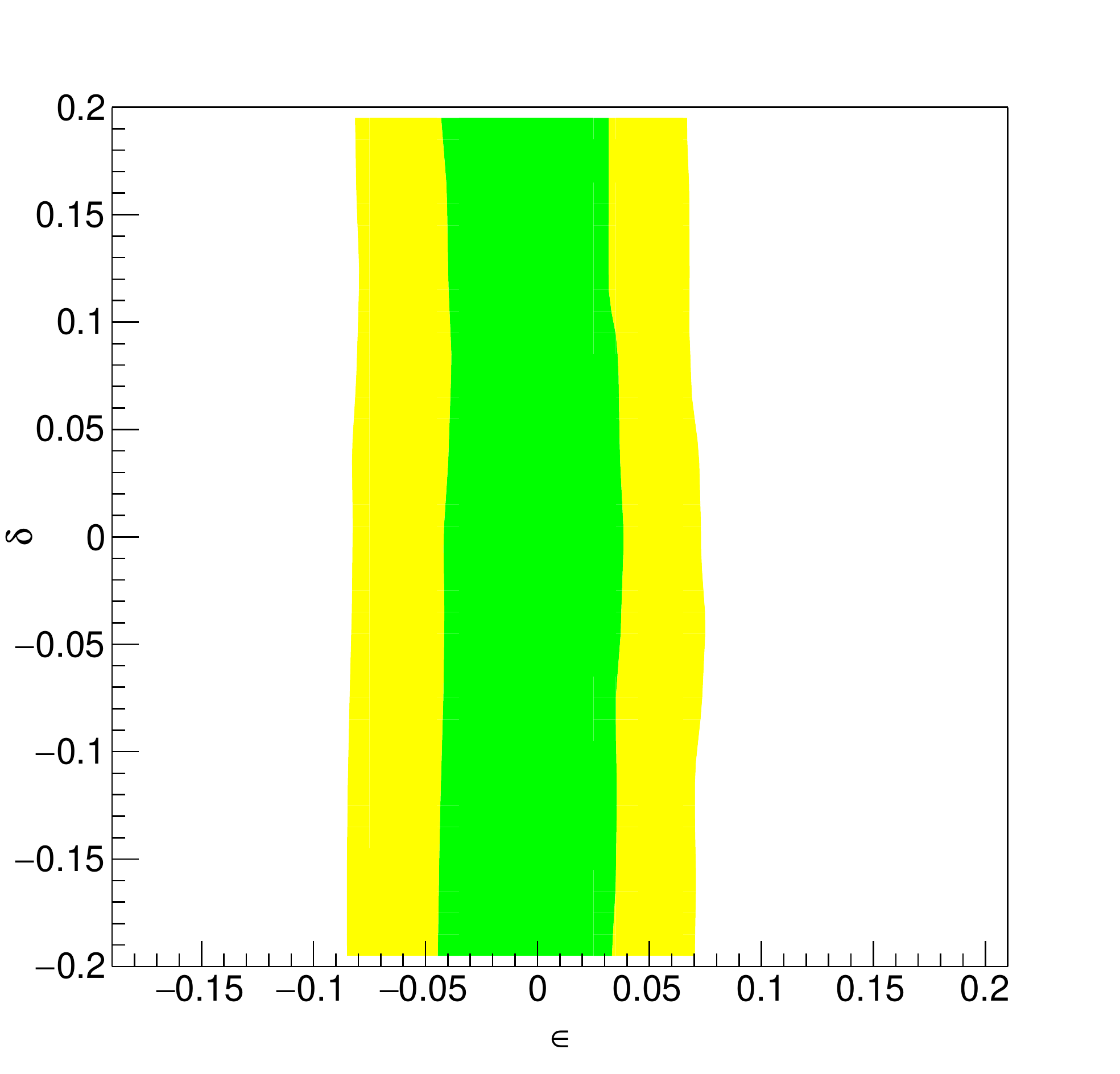} }}%
    \caption{Constraints on  the $\epsilon $ and $\delta$ parameters for $28$ TeV collision energy, $n=k=15$ boundary for DR, SR and NR regions. Green and yellow arias correspond to exclusion limits  at $68\%$ and $95\%$ CL on $\epsilon $ and $\delta$ for the CS measured with $8\%$ uncertainty.}%
    \label{fig:28_TeV_limits}%
\end{figure}

\begin{figure}%
    \centering
    \subfigure[~DR region]{{\includegraphics[width=6cm]{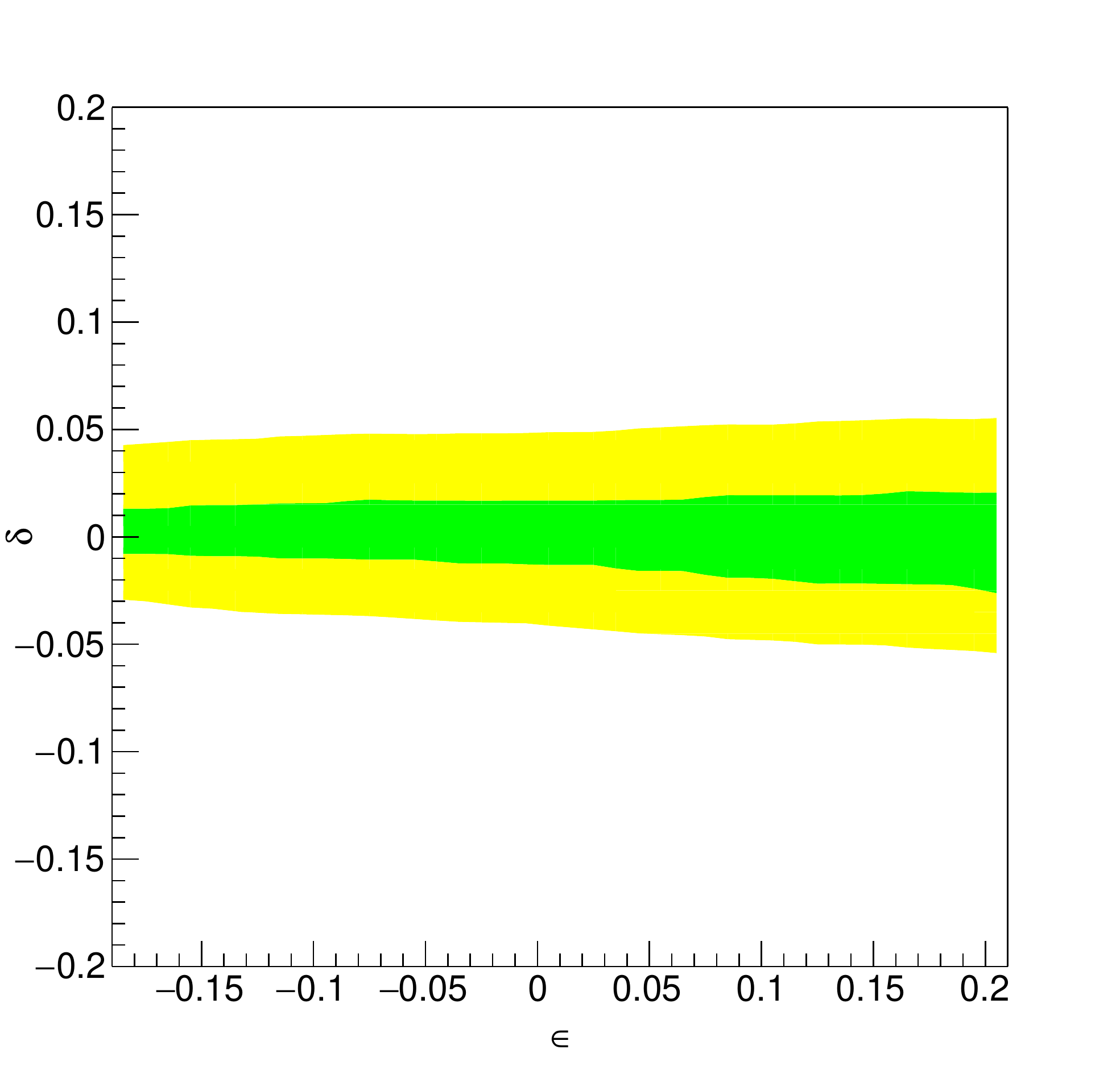} }}%
		\qquad
    \subfigure[~SR region]{{\includegraphics[width=6cm]{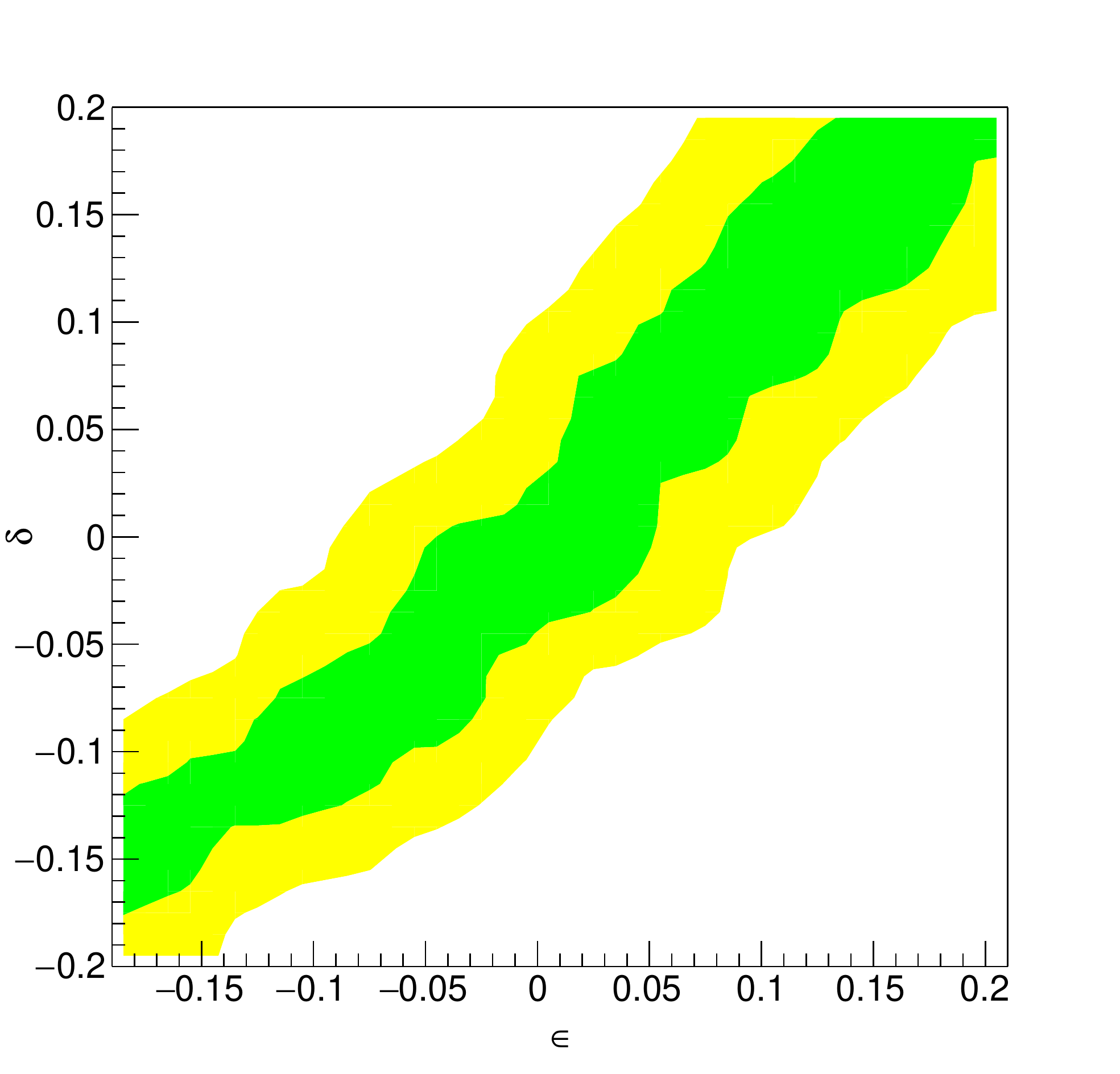} }}%
    \qquad
    \subfigure[~NR region]{{\includegraphics[width=6cm]{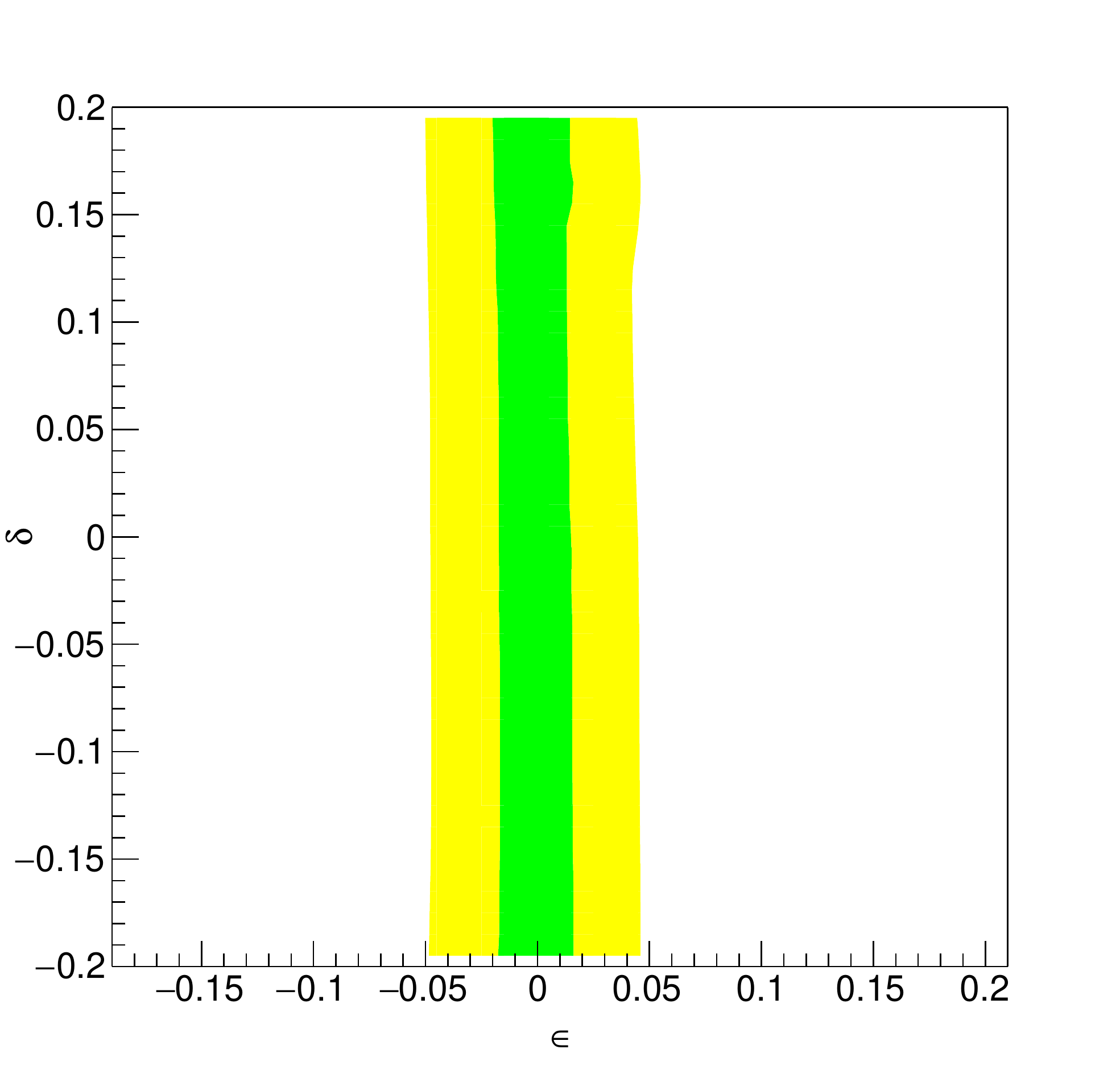} }}%
    \caption{Constraints on  the $\epsilon $ and $\delta$ parameters for $100$ TeV collision energy, $n=k=15$ boundary for DR, SR and NR regions. Green and yellow arias correspond to exclusion limits  at $68\%$ and $95\%$ CL on $\epsilon $ and $\delta$ for the CS measured with $5\%$ uncertainty.}%
    \label{fig:100_TeV_limits}%
\end{figure}

For the fitting, we considered the cross section obtained in each region as a random variable depending on two parameters $\epsilon$ and $\delta$. By choosing the appropriate $\chi^2$ distribution quantiles of $2.3$ and $6$ for 68 and 95\% confidence level correspondingly we derive upper limits on $\epsilon $ and $\delta$  Fig.~\ref{fig:14_TeV_limits}. Similar plots for $28$ TeV and $100$ TeV  are also obtained Figs.~\ref{fig:28_TeV_limits},~\ref{fig:100_TeV_limits}. 

\begin{figure}%
    \centering
    \subfigure[~$14$ TeV]{{\includegraphics[width=6cm]{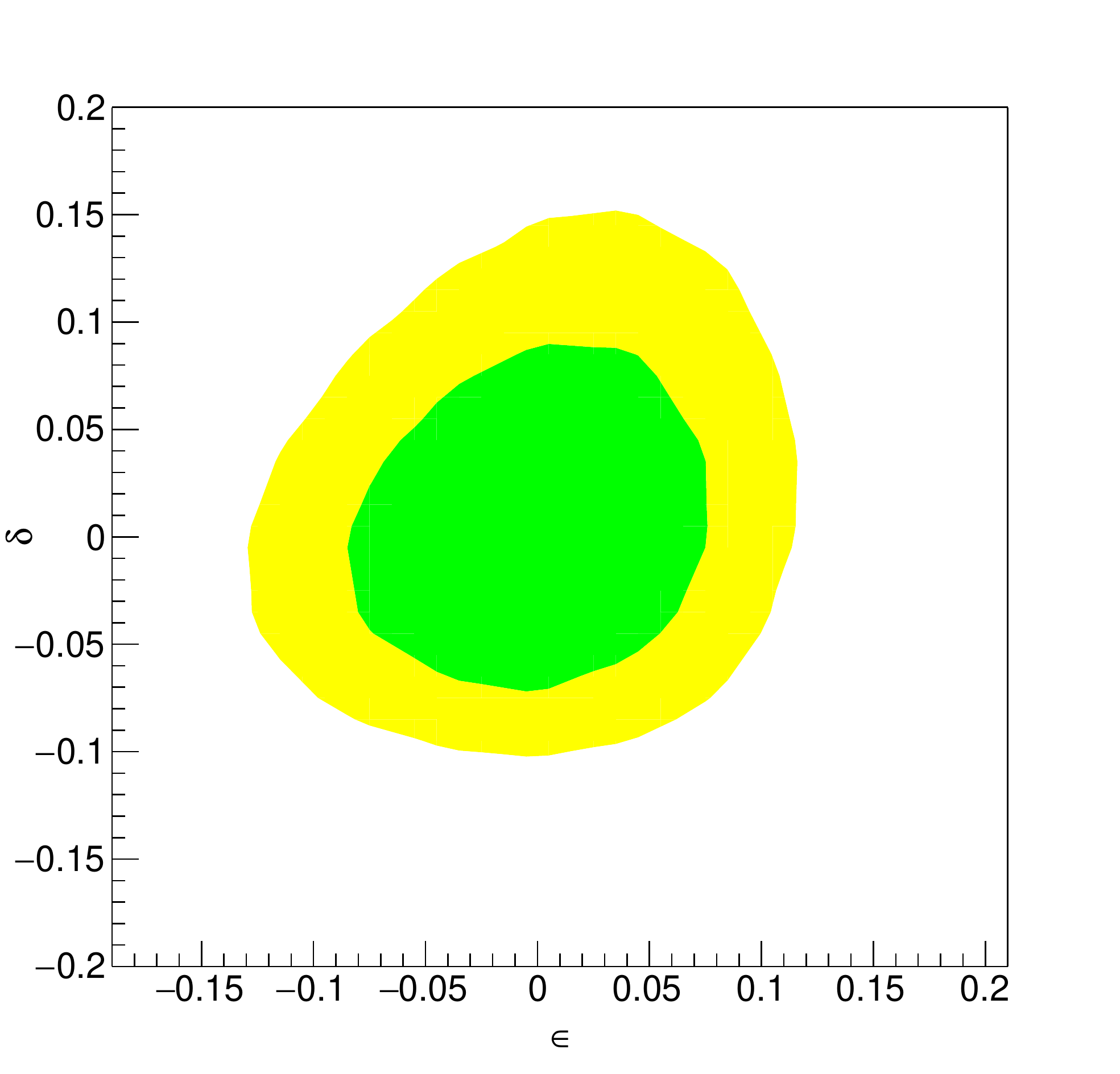} }}%
		\qquad
    \subfigure[~$28$ TeV]{{\includegraphics[width=6cm]{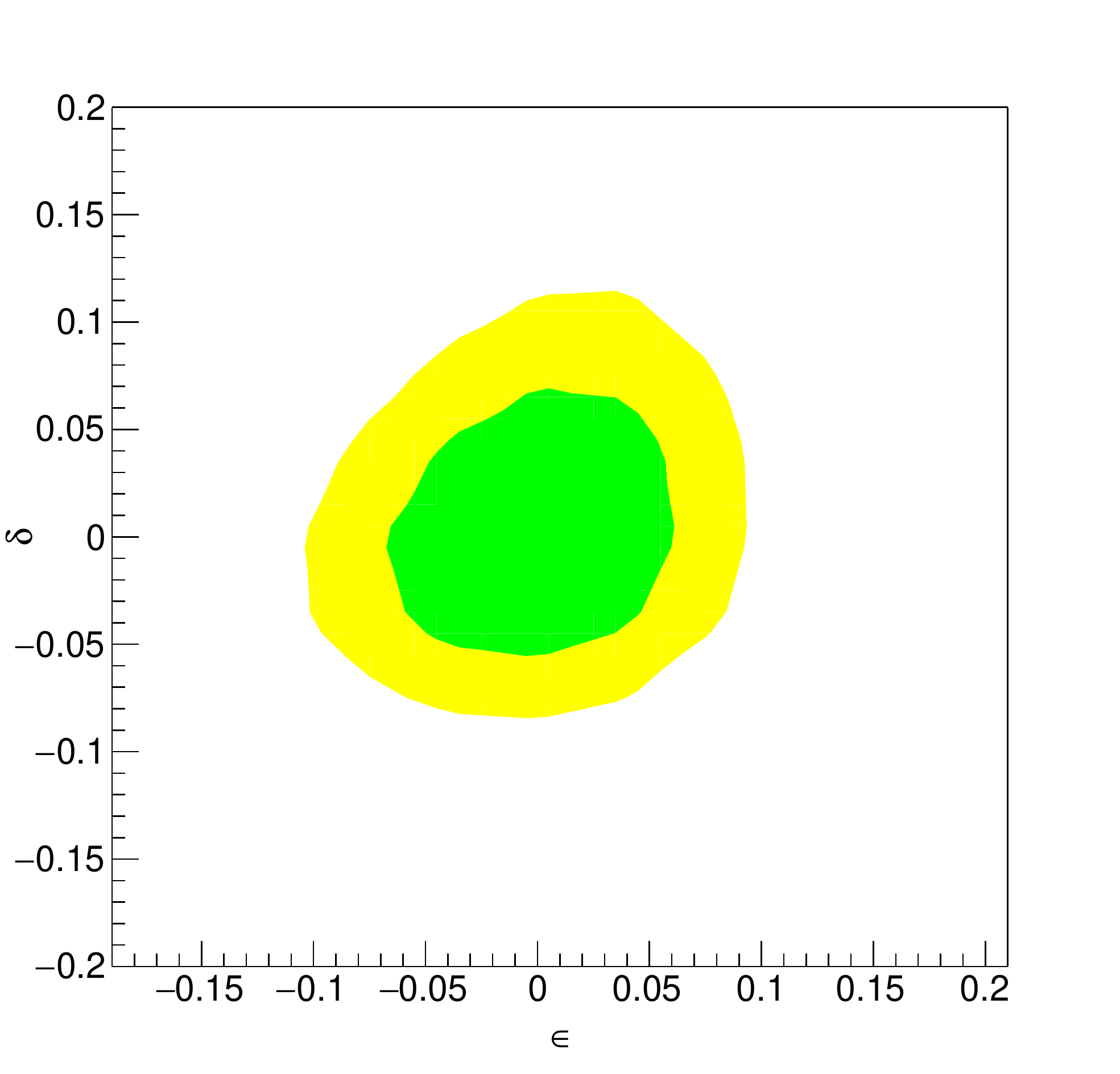} }}%
    \qquad
    \subfigure[~$100$ TeV]{{\includegraphics[width=6cm]{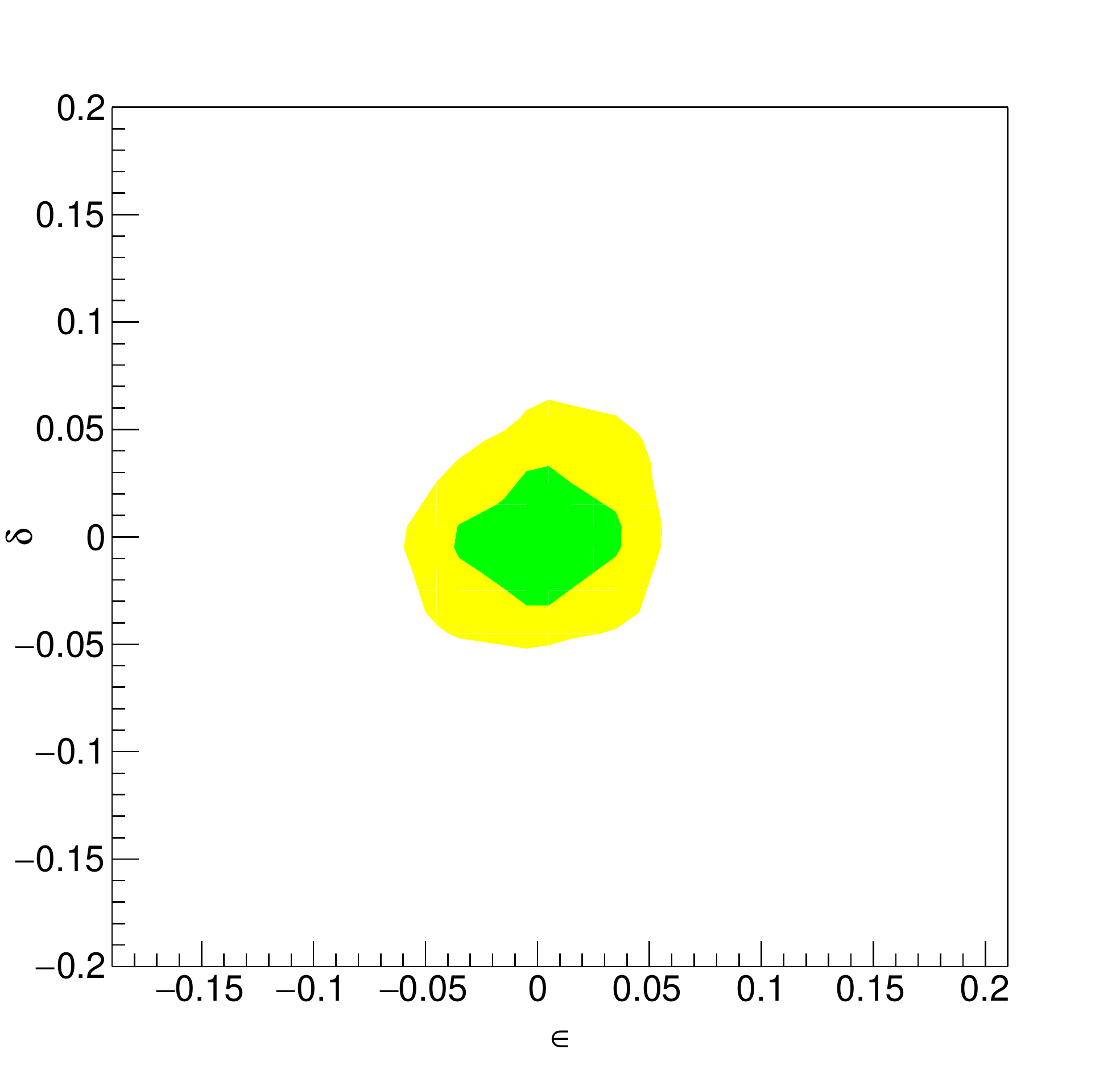} }}%
    \caption{Combined constraints on  the $\epsilon $ and $\delta$ parameters for DR, SR and NR regions for different collision energies. Green and yellow arias correspond to exclusion limits  at $68\%$ and $95\%$ CL on $\epsilon $ and $\delta$.}%
    \label{fig:combined_limits}%
\end{figure}

As events in DR, SR and NR regions do not overlap Eq.~\ref{eq1}, for combination fitting results were summed in terms of $\chi^2$ with an increase of appropriate quantiles for six d.o.f. $7$ and $12.6$ Fig.~\ref{fig:combined_limits}.  From the limits on the parameters, $\epsilon$ and $\delta$ one gets achievable constraints on the top quark width using  Eq.~\ref{eq3} and combining restrictions on the $\epsilon$ and $\delta$ parameters in quadratures. Model-independent constraints on the top quark width are estimated to be $23\%$ and $12\%$ for the energies 14 to 100 TeV respectively with assumed experimental accuracy of fiducial cross section measurements to be $10\%$ and $5\%$ Table.~\ref{tab1}.

\begin{table}[]
\begin{tabular}{l|l|l|l}
 $ \sqrt{s} $ in TeV  & $14$  & $28$  & $100$   \\ \hline
  given experimental uncertainty in percentages   & 10 & 8  & 5 \\ \hline
  $\Gamma_t$ shift in GeV & 1.15 - 1.83 & 1.21 - 1.77 & 1.31 - 1.67 \\ 
\end{tabular}
\caption{Estimations of the top quark width shift in case of the DR, SR and NR regions cross section measured  with a given experimental uncertainty.}
\label{tab1}
\end{table}

\section{Summary and outlook}
\label{Summary}

Gauge invariant estimation of deviations of the top quark width from the SM value is obtained in different kinematic regions. It is shown that top quark production cross section in the double-resonant region is most sensitive to the $\delta$ parameter, which modifies only the top quark width. The fiducial cross section in the non-resonant region has the sensitivity to $\epsilon$ parameter through a modification the Wtb coupling in the amplitude numerator. The single-resonant region has a comparative sensitivity to both parameters since the $\delta$ parameter modifies the top quark width and the $\epsilon$ parameter modifies both the top quark width and the Wtb coupling. 
The significant difference in dependence of fiducial cross sections in DR, SR and NR regions on $\epsilon$ and $\delta$ parameters one of the main observation of this study.
This fact allows to put combined limits on $\delta$ and $\epsilon$  parameters simultaneously and using these limits to obtain constraints on the top quark width. Achievable constraints in the model-independent way on the top quark width are estimated to be from $23\%$ to $12\%$ for corresponding experimental accuracy from $10\%$ to $5\%$. These results are obtained using a simplified approach when all inaccuracies are encoded into assumed overall systematic uncertainty. Detail study of all effects such as hadronization and fragmentation, detector response as well as an impact of backgrounds are beyond the scope of the current simplified study when only the main idea is demonstrated. Detalisation of abovementioned effects is postponed to the next more realistic analysis.

\section{Acknowledgments}
 The work was supported by grant 16-12-10280 of Russian Science Foundation.

\end{document}